\documentclass{iau}
\usepackage{graphicx}
\usepackage{hyperref}

\title[Stochastically excited oscillations on the upper main sequence] 
{Stochastically excited oscillations on the upper main sequence}

\author[V. Antoci]   
{Victoria Antoci}

\affiliation{Stellar Astrophysics Centre, Department of Physics and Astronomy, Aarhus University \\ 
Ny Munkegade 120, DK-8000 Aarhus C, Denmark\\ 
email: {\tt antoci@phys.au.dk \\[\affilskip]}}

\pubyear{2014}
\volume{301}  
\pagerange{1--8}
\jname{Precision Asteroseismology}
\editors{J.A. Guzik, W.J. Chaplin, G. Handler \& A. Pigulski, eds.}
\begin{document}

\maketitle

\begin{abstract}
Convective envelopes in stars on the main sequence are usually connected only with stars  of spectral types F5 or later. However, observations as well as theory indicate that the convective outer layers in earlier stars, despite being shallow, are still effective and turbulent enough to stochastically excite oscillations. Because of the low amplitudes, exploring stochastically excited pulsations became possible only with space missions such as {\it Kepler} and \textit{CoRoT}. 
Here I review the recent results and discuss among others, pulsators such as $\delta$ Scuti, $\gamma$ Doradus, roAp, $\beta$ Cephei, Slowly Pulsating B and Be stars, all in the context of solar-like oscillations.

\end{abstract}

\firstsection 
\section{Introduction}

 There are several mechanisms driving oscillations in pulsating variables. The oldest known is the opacity ($\kappa$) mechanism acting like a heat engine, converting thermal energy into mechanical energy (e.g. Eddington 1919, Cox 1963). The layers where the thermal energy is stored are the zones connected to (partial) ionisation of abundant elements, which can take place only at specific temperatures. The zone where neutral hydrogen (H) and helium (He) are ionised is at about 14\,000~K and close to the surface, whereas the second ionisation zone of He II is at $\sim$50\,000~K. The driving in the He II ionisation zone is the main source of  excitation in stars placed in the classical instability strip such as the $\delta$ Scuti ($\delta$ Sct) stars. The pulsations of the more massive $\beta$ Cephei ($\beta$~Cep) and Slowly Pulsating B stars (SPB) are triggered by the $\kappa$ mechanism operating on the iron-group elements (located at $\sim$200\,000~K). 
 
\begin{figure}[]
\begin{center}
 \includegraphics[width=8cm]{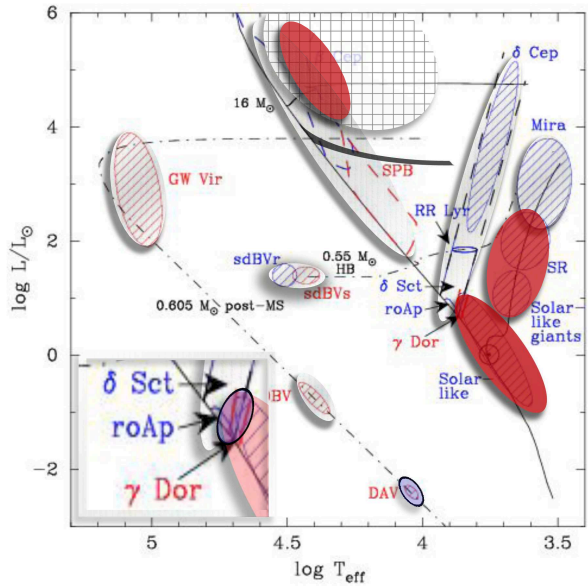} 
 \caption{The asteroseismic H-R diagram displays the major types of  known pulsating stars. This modified version shows their excitation mechanisms labeled as follows: the $\kappa$ mechanism is depicted in light grey, the stochastic mechanism in dark grey (red in the electronic form) and the grey area surrounded by a black ellipse (blue area in the electronic form) indicates the convective blocking mechanism. The inset is a close-up of the lower part of the classical instability strip. Note the overlap of the different instability strips. The upper hashed region depicts the region where massive stars have a sub-surface convection zone with convective velocities $\ge$ 2.5~km\,s$^{-1}$. The black line, just above the SPB stars shows the division line for surface convection in the Fe ionisation zone. Both regions indicating convection in massive stars are from Cantiello et al.~(2009). Adapted from J. Christensen-Dalsgaard and G. Handler.  }
   \label{hrd}
\end{center}
\end{figure}

 The oscillations in $\gamma$ Doradus stars ($\gamma$ Dor) are driven by a similar mechanism, in this case, however, it is the bottom of the outer convection zone blocking the flux from the interior (Guzik et al.~2000) and is therefore called {\it convective blocking } (see Fig. \ref{hrd} for exact location in the Hertzsprung-Russell diagram).

Another important mechanism, especially for the present review, is
stochastic driving. In stars like the Sun the modes are intrinsically
stable, damped by the turbulent convection. Nevertheless the acoustic
energy of the convective motion, which is comparable to the sound
speed, is sufficient to cause resonance at a star's natural
frequencies where a part of the energy is transferred into global
oscillation modes. Because of the large number of convective cells the
excitation is random, hence stochastic. This is very different from
the $\kappa$ mechanism, which excites pulsation coherently. This
contrast is a very important way of distinguishing between these two
types of driving mechanisms in the signal processing. Another
important attribute of stochastic driving is that all the modes in a
certain frequency range are excited to observable amplitudes, allowing
mode identification from pattern recognition. This is again in
absolute contrast to the heat engine where the mechanism selecting
which modes are excited to observable amplitudes is not understood (for
a review see Smolec, these proceedings).

However, a star will only pulsate if the conditions inside are just right
for a mechanism to work. For the $\kappa$ mechanism, a mode will only
be excited if its period of oscillation corresponds to the thermal
time scale of the driving zone (e.g. Pamyatnykh 2000), which means
that the opacity bump needs to be at a favourable depth in the
envelope. Similarly for the convective blocking mechanism, the depth of
the convective envelope should be between 3 and 9\% (Guzik et
al.~2000). As for the stochastic excitation, what matters is not the
depth of the convective layer, but how vigorous the convective motions
are.

Stochastically excited pressure modes as observed in the Sun are
nearly equally spaced in frequency and give rise to a clear comb-like
structure. The almost equidistant spacing in frequency of consecutive
modes of the same degree $\ell$ is the so called {\it large frequency
  separation} $\Delta\nu$ and measures the mean density of a
star.  The
oscillation spectrum in Sun-like stars as well as in red giants is
described by an envelope with a frequency of maximum power, for the
Sun $\nu_{\rm max}$ is at 3090 $\mu$Hz (Huber et al.~2011). The
$\nu_{\rm max}$ and the shape of the oscillation envelope, determined
by damping and driving, are related to the acoustic
cutoff frequency, $\nu_{\rm ac}\propto g/\sqrt{T_{\rm eff}}$
(e.g. Brown et al.~1991). From that, Kjeldsen \& Bedding (1995)
derived the following scaling relation for $\nu_{\rm max}$ for any
other star: $\nu_{\rm max}=\nu_{\rm
  max_{\odot}}({M/M_{\odot}})/[(R/R_\odot)^2 \sqrt{T_{\rm eff}/T_{\rm
      eff_{\odot}}}]$. Kjeldsen \& Bedding (1995) also show that
$\Delta\nu$, when scaled to the Sun, can also be written as
$\Delta\nu=(M/M_{\odot})^{1/2}(R/R_{\odot})^{-3/2}
\Delta\nu_{\odot}$. From observations with the \textit{CoRoT} and {\it
  Kepler} satellites it was demonstrated that there is a clear
empirical relation between $\nu_{\rm max}$ and $\Delta\nu$ of the
form: $\Delta\nu=\alpha(\nu_{\rm max}/\mu {\rm Hz})^\beta$, where
$\alpha$ and $\beta$ can have slightly different values, summarised by
e.g.~Huber et al.~(2011). All these relations are powerful tools to
study solar-like oscillators.

The timescale on which a solar-like mode is excited, the mode lifetime,  depends on the damping rate. This can be determined by fitting a Lorentzian profile to an observed mode and measuring its line width, where the inverse of the FWHM times $\pi$ gives the mode lifetime. Because the power of the mode is spread over a certain  frequency range the amplitudes and frequencies are also derived by fitting Lorentzian profiles.

Detecting solar-like oscillations in Sun-like stars is easy, however identifying stochastically excited modes in other types of pulsators is complicated and requires a lot of data. In particular discriminating between different excitation mechanisms is sometimes impossible. Temporal variability does not necessarily mean stochastic excitation; it can simply be that there are many closely spaced, unresolved frequencies. The good news is that even in the case where coherent and non-coherent signals are present, one can subtract the coherent peaks without destroying the stochastic signal (Antoci et al.~2013).

Studying the excitation mechanisms of stars with different temperatures, masses and evolutionary stages allows us to determine the structure of stars as well as understand several physical processes. The existence of convective layers plays a major role in the generation of magnetic fields and stellar activity,  in the transport of angular momentum, in mixing processes and diffusion, and even in the alignment of planetary systems.

\section{A and early F type stars}

\begin{figure}[t]
\begin{center}
 \includegraphics[width=10cm]{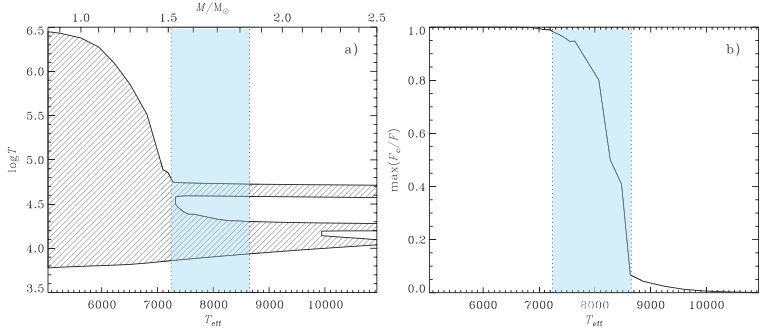} 
 \caption{Envelope structure for zero-age main sequence stars as a function of effective temperature (lower abscissa) and mass (upper abscissa). \textit{Left:} the dashed area indicates the convective regions in the outer layers. It can be seen that the depth of the convective envelope changes dramatically within the classical instability strip (grey area, blue in electronic form). While for stars like the Sun, all three convection zones are combined, for hotter more massive stars the H, He I and He II zones start to separate. \textit{Right:} $F_{\rm c}/F$ indicates how much of the energy is transported by convection. Again a clear transition is present within the instability strip. Courtesy of J.~Christensen-Dalsgaard.}
   \label{conv}
\end{center}
\end{figure}

The A and early F type stars are very complex objects, because the
transition from deep and effective to shallow convective envelopes
takes place within this spectral region. Figure \ref{conv} shows how the
outer layers change as a function of temperature for stars located on
the zero-age main sequence, from theory (Christensen-Dalsgaard
2000). While the outer 30\% is fully convective for the Sun, it
changes dramatically for slightly higher temperatures and separates
into the known ionisation zones (H, He I and He II). For a late A type
star, for example, only the outer 2\% of the radius is still
convective (e.g. Kallinger \& Matthews 2009). The right panel of
Fig.~\ref{conv} shows how much of the energy is transported by
convection, demonstrating again the drastic change in the convective
properties. The grey (electronic version blue) region in the same
figure indicates the location where the $\kappa$ mechanism operates in
the He II ionisation zone, as assumed to be the case in the $\delta$
Sct stars. Interestingly, Th\'eado et al.~(2012) find that if diffusion
is implemented in the models, heavy elements will gravitationally
settle. Due to the strong accumulation of Fe in the Z bump, this
region eventually becomes convective, mixing again the material into
the higher layers connected by overshooting, hence destroying the
steep gradient. Once this layer is homogeneous enough the convection
will cease and the cycle repeats. The authors find several convective
episodes during the main sequence lifetime of a 2~$M_{\odot}$ star.

From an observational point of view, Gray \& Nagel (1989) found from
bisector measurements that the granulation boundary crosses the
classical instability strip. Also related to the
afore-mentioned transition is the so-called rotation boundary. Stars
later than F5 have slow rotation rates, on average $v\sin i < $ 10
km\,s$^{-1}$, with a sharp increase between F0 and F5 (Royer
2009). The measured rotational velocities for the cool stars inside
the instability strip are typically higher than 100
km\,s$^{-1}$. Royer (2009) also reports observational evidence for
differential rotation in A type stars, which is related to magnetic
dynamos like in the Sun. Chromospheric activity disappears only for
stars hotter than 8300~K (Simon et al.~2002), demonstrating that the
$\delta$ Sct and $\gamma$ Dor pulsators are still affected by the
presence of convective envelopes. Landstreet et al.~(2009) and
Kallinger \& Matthews (2009) find strong evidence for the signatures
of convective motions.

Houdek et al.~(1999) and Samadi et al.~(2002) predicted that surface
convection in the envelopes of $\delta$ Sct stars is still vigorous
enough to excite solar-like oscillations. There is no reason why the
same prediction should not be generalised for all the stars occupying
the same location in the H-R diagram. Solar-like oscillations, if detected,
would allow one to perform mode identification from pattern
recognition and therefore to also identify the modes of pulsation
excited by the $\kappa$ mechanism. As a result we can then reproduce
the deep interior of the star by using the $\delta$ Sct oscillations
and constrain the overshooting parameter $\alpha_{\rm ov}$, which is
the primary factor to influence the lifetime on the main sequence. The
solar-like oscillations on the other hand are most sensitive to outer
(convective) envelope, which so far is poorly understood in stars with
temperatures higher than 6500~K.

\subsection{(Rapidly oscillating) Ap stars}

Rapidly oscillating (ro)Ap stars are stars with magnetic fields of the
order of kG, with periods of oscillations similar to the solar ones (4
to 20 min).  The regular patterns in the oscillation spectrum of roAp
stars are best explained by the oblique pulsator model, where the
pulsation axes is inclined with respect to the magnetic and rotation
axes (Kurtz 1982, Bigot \& Dziembowski 2002). The excitation mechanism
is not entirely understood but is interpreted as the opacity mechanism
acting in the H ionisation zone. In Ap stars the magnetic field
suppresses convection at the magnetic poles reducing the damping
(Balmforth et al.~2001), triggering the observed p modes. So far there
are no solar-like oscillations observed in roAp stars, but also none
expected.

\subsection{$\gamma$ Dor stars}

 The generally accepted mechanism triggering the high radial order g
 mode pulsations in $\gamma$ Dor stars is the convective blocking
 mechanism. Interestingly, the instability domain of the $\gamma$ Dor
 stars overlaps the one of the $\delta$ Sct stars, suggesting that
 both types of pulsation should be excited simultaneously in stars
 located there (Fig. \ref{hrd}). Missions like the NASA {\it Kepler}
 satellite established that the $\delta$ Sct and $\gamma$ Dor hybrids
 are a common phenomenon, rather than an exception (Grigahc\`{e}ne et
 al.~2010).\ Intriguingly, from {\it Kepler} observations,
 Uytterhoeven et al.~(2011) and Tkachenko et al.~(2012) find hybrid
 pulsators not only in the region where the two instability domains
 overlap but distributed all over the $\delta$ Sct strip. The
 convective blocking mechanism, as it is currently understood, cannot
 operate unless the convective envelope has a certain depth between
 approximately 3 and 9\% (Guzik et al.~2000); this is not expected to
 be the case at the observed effective temperatures. This implies that
 either the hybrids observed at such high $T_{\rm eff}$ values have
 still a non-negligible subsurface convection zone or we are missing
 basic physics.  As far as stochastic excitation is concerned, these
 stars are the perfect targets to search for solar-like
 oscillations. Nevertheless, up to now, no $\gamma$ Dor/solar-like
 hybrids were detected, which is very puzzling. The non-detection of
 stochastic oscillations in these stars becomes even more mysterious
 when considering the significant number of solar-like oscillators
 identified within the $\gamma$ Dor instability strip. In other words,
 there are several F type stars with high temperatures, higher than
 some of the $\gamma$ Dor pulsators, showing solar-like oscillations
 but no g modes. Note that here I consider only targets with reliable
 temperature determination (W.~J.~Chaplin, priv. comm.). To summarise,
 there are no solar-like oscillations detected in $\gamma$ Dor stars,
 but from theoretical point of view we do not understand why. One has
 to keep in mind, however, that $\gamma$ Dor are usually observed with
 a sampling cadence not rapid enough to detect high radial order p
 modes.

\subsection{$\delta$ Sct stars}

$\delta$ Sct stars are one of the first known groups of pulsators, yet
also one of the least understood. It is generally accepted that the
complex pulsational behaviour is the result of the $\kappa$ mechanism
acting in the He II ionisation zone.  Previously, models predicted
more modes to be unstable than observations showed; however, with
\textit{CoRoT} and {\it Kepler}, the opposite seems to be the case
(e.g.~Poretti et al.~2009, Garc\'{\i}a-Hern\'andez et al.~2010, Balona
\& Dziembowski 2011, Uytterhoeven et al.~2011). The NASA {\it Kepler}
spacecraft observed hundreds of $\delta$ Sct stars at a precision
where solar-like oscillations, if present, should be detected. One of
these stars was HD~187547 (Antoci et al.~2011), at first glance a
typical $\delta$ Sct star. However, at high frequencies, the authors
detected modes approximately equidistantly spaced, as expected for
high radial order p-modes, which are not combination frequencies. The
$\kappa$ mechanism, as it is known, cannot excite a continuous
frequency region as observed in HD~187547 (Pamyatnykh
2000). Spectroscopic observations exclude that the peaks at high
frequencies originate from a possible companion; such a star would be
an A or F-type star and would be visible in the spectrum, which is
not. The measured large separation $\Delta \nu$ predicts $\nu_{\rm
  max}$ to be exactly where the mode with the highest amplitude in the
supposed stochastic region is detected. New data, however showed that
the mode lifetimes are longer than the observing run, suggesting
coherence over more than 450 days (from scaling relations the mode
lifetimes are suggested to decrease with increasing $T_{\rm
  eff}$). This is not compatible with the interpretation of `pure'
solar-like oscillations. The word `pure' in this context should
emphasise that there are no studies on how the $\kappa$ mechanism
interacts with the stochastic excitation, especially in the case where
the periods of pulsations have similar time scales.

Very recent results, in collaboration with Margarida Cunha and G\"unter Houdek, suggest the presence of a new excitation mechanism for at least the $T_{\rm eff}$ domain of HD~1787547, i.e.~7500~K. The models, using a non-local and time-dependent treatment of convection (Houdek et al.~1999 and references therein), reproduce the major part of the observed modes in HD~187547, as being excited by the turbulent pressure and not the $\kappa$ mechanism. The turbulent pressure can be understood as the dynamical component of convection, associated with the transport of momentum (details will be published in a dedicated article). This proves that convection in the envelope of these  stars still plays an important role and can excite pulsations even if not stochastically.

\section{O and B type stars}

When it comes to stellar structure, O and B type stars are usually described as having a convective core and a purely radiative envelope. Cantiello et al.~(2009), however, show that this is not necessarily true (Fig.~\ref{hrd}). Even though it might comprise a negligible part of the stellar mass, subsurface convection occurring predominantly in the Z bump can not only lead to variability in the observed microturbulent velocity but also influence the stellar evolution of massive stars significantly. This is because the photospheric motion caused by convection can strongly influence mass loss by affecting the stellar winds known to exist in these stars (see Cantiello et al.~2009). Furthermore, a convection zone close to the surface can also induce magnetic fields, which may favour mass loss and more importantly, loss of angular momentum.

Samadi et al.~(2010) investigated from a theoretical point of view
whether the turbulent convection can stochastically excite g modes in
massive stars. They find that, while low radial order g modes might be
excited by the convection in the core, the convective envelope is
primarily exciting high radial order g modes. However, the same
authors predict the amplitudes to be too low to be observed even with
\textit{CoRoT} or {\it Kepler}. Belkacem et al.~(2010), on the other
hand, find in their exploratory models of a 10~$M_{\odot}$ that both
the core and the convective outer layers associated with the Fe bump
can efficiently drive stochastic oscillations with detectable
amplitudes. However, the authors also comment that the convective
properties in these temperature domains are poorly understood. Shiode
et al.~(2013), arrive to a similar conclusion, namely that the
convection in the core can stochastically excite g modes to observable
amplitudes. They expect the amplitudes to be highest for stars with
masses $\ge$ 5~$M_{\odot}$. Also from 3D MHD and numerical
simulations Browning et al.~(2004) and Rogers et al.~(2012) find that
convection in the core excites internal gravity modes in stars more
massive than the Sun.  From the observational point of view there is
strong evidence for the presence of stochastically excited
oscillations in massive O type stars observed with
\textit{CoRoT}. HD~46149 (O8.5\,V) appears to oscillate in stochastic
modes fulfilling the scaling laws as expected for solar-type stars and
also showing ridges for modes of equal degree but consecutive radial
orders as those observed for Sun-like stars (Degroote et
al.~2010). The $\kappa$ mechanism is not expected to drive the
observed oscillations. Besides that, the temporal variability argues
in favour of their stochastic nature. Blomme et al.~(2011) find 300
significant frequencies in HD 46966 (O8\,V), all displaying temporal
variability which might be due to granulation noise. The same authors
find similar behaviour in other two O type stars.

\subsection{SPB stars}

SPB stars pulsate in high radial order g modes excited by the $\kappa$ mechanism associated with the opacity in the Z bump (Fig.~\ref{hrd}). Their instability region overlaps that of the $\beta$~Cep stars, suggesting the presence of both type of pulsations. As in the case of $\gamma$ Dor and $\delta$ Sct stars, observations are consistent with theory (e.g. Handler et al.~2009). According to Cantiello et al.~(2009), the only subsurface convective layer present in these stars is in the He II ionisation zone, but  it is shallow and inefficient. No solar-like oscillations driven by the convection in the envelope are expected in SPB stars. If there would be any, these would be excited by the convection in the core but none were observed so far. 

\subsection{$\beta$ Cep stars}
The $\beta$ Cep stars pulsate in p and g modes excited by the $\kappa$ mechanism in the Z bump. From Cantiello et al.~(2009) and the theoretical work summarised above, these stars are expected to have a substantial convection layer in the Z bump. Belkacem et al.~(2009) reported the detection of solar-like oscillations and opacity driven modes in the $\beta$ Cep star V1449 Aql. The authors argue that the peaks observed at high frequencies are consistent with stochastic oscillations, because of the temporal variability and  the detected spacing interpreted as the large frequency separation. Aerts et al.~(2011), on the other hand, can seismically reproduce the pulsational behaviour of V1449 Aql without invoking the presence of solar-like oscillations but only $\kappa$ mechanism excitation. They subscribe the power at higher frequencies to non-linear resonant mode coupling between the dominant radial fundamental mode and many other low-order p modes. Degroote (2013) delivers a third possible explanation for the observed frequency spectrum of V1449 Aql, arguing that chaotic behaviour of the very dominant mode can qualitatively reproduce the observed frequency spectrum. As no other member of the $\beta$ Cep pulsators was found to show solar-like oscillations, the case of V1449 Aql is still a matter of debate.

\subsection{Be stars}

Be stars are stars showing emission in their spectra originating from a circumstellar disk. These can be found spreading over several spectral classes (O, B and early A). So far the Be phenomenon is still not fully understood but can be related to rapid rotation, and meanwhile to pulsations (of SPB and $\beta$~Cep type); the latter were found to be present in all Be stars examined so far (Neiner et al.~2013). Other types of variability attributed to rotation, magnetic fields, stellar winds and outbursts were also observed in these stars (for a complete review on Be stars see  Porter \& Rivinius 2003). A very fortunate case is the Be star HD\,49330 (B0\,IVe) which was observed during an outburst with \textit{CoRoT} as well as from ground (Huat et al.~2009). In this star the authors find a clear correlation between the outburst and the presence of g modes. While during the quiescent phase the pulsations are predominantly in the p mode regime, during the outburst the g modes are enhanced to higher amplitudes and are also very unstable indicating short mode lifetimes. The modelling of this star, even considering rapid rotation, shows that the $\kappa$ mechanism cannot excite the observed g modes. A very recent study of another Be star (HD\,51452, Neiner et al.~2012) suggests that the g modes observed in the two afore-mentioned Be stars are in fact stochastically driven gravito-inertial modes. These oscillations are excited by the convection in the core and enhanced to observable amplitudes by the rapid rotation (see Neiner et al.~2012 and Mathis et al.~(2013) for details).

\section{Conclusions}

The potential to detect stochastically excited oscillations in stars
significantly more massive than the Sun became possible only with the
space missions \textit{CoRoT} and {\it Kepler}. It is not only the
quality of the data which is outstanding, but also the uninterrupted
and long observing seasons which demonstrated that we need to revisit
our knowledge on pulsation mechanisms (Fig.~\ref{hrd}). Convection is
one of the most complex 
processes in astrophysics and therefore often neglected with the
argument that convection zones in stars hotter than early F are
negligible. However, the latest results reviewed in this article
demonstrate that the impact of convection needs to be taken into
account.

\begin{acknowledgements}
Funding for the Stellar Astrophysics Centre (SAC) is provided by The Danish National Research Foundation. The research is supported by the ASTERISK project (ASTERoseismic Investigations with SONG and {\it Kepler}) funded by the European Research Council (Grant agreement no.: 267864). 

\end{acknowledgements}



\begin{thebibliography}{}

\bibitem[Aerts et al.~(2011)]{ }
{ Aerts, C., Briquet, M., Degroote, P., Thoul, A., \& van Hoolst, T.} 2011,
\textit{A\&A}, 534, A98

\bibitem[Antoci et al.~(2011)]{ }
{Antoci, V., Handler, G., Campante, T.L., et al.} 2011, 
\textit{Nature}, 477, 570

\bibitem[Antoci et al.~(2013)]{ }
{Antoci, V., Handler, G., Grundahl, F., et al.} 2013, \textit{MNRAS}, in press

\bibitem[Balmforth et al.~(2001)]{ }
{Balmforth N.J., Cunha, M.S., Dolez, N., Gough, D.O., \& Vauclair, S.} 2001, \textit{MNRAS}, 323, 362

\bibitem[Balona \& Dziembowski (2011)]{ }
 {Balona, L.A., \& Dziembowski, W.A.} 2011, \textit{MNRAS}, 417, 591

\bibitem[Belkacem et al.~(2009) ]{ }
{Belkacem, K., Samadi, R., Goupil, M.-J., et al.} 2009, \textit{Science}, 324, 1540 

\bibitem[Belkacem et al.~(2010)]{ }
{Belkacem, K., Dupret, M.-A., \& Noels, A.} 2010, \textit{A\&A}, 510, A6
 
\bibitem[Bigot \& Dziembowski (2002)]{ }
{Bigot, L., \& Dziembowski, W.A.} 2002,  \textit{A\&A}, 391, 235

\bibitem[Blomme et al.~(2011)]{ }
{Blomme, R., Mahy, L., Catala, C., et al.} 2011, \textit{A\&A}, 533, A4

\bibitem[Brown et al.~(1991)]{ }
{Brown, T.M., Gilliland, R.L., Noyes, R.W., \& Ramsey, L.W.} 1991, \textit{ApJ}, 368, 599

\bibitem[Browning et al.~(2004)]{ }
{Browning, M.K., Brun, A.S., \& Toomre, J.} 2004, in: J. Zverko, J. \v{Z}i\v{z}\v{n}ovsk\'y, S.J. Adelman, \& W.W. Weiss (eds.), \textit{The A-Star Puzzle}, Proc.~IAU Symposium No.~224 (Cambridge: Cambridge University Press), p.\ 149

\bibitem[Cantiello et al.~(2009)]{ }
{Cantiello, M., Langer, N., Brott, I., et al.} 2009, \textit{A\&A}, 499, 279

\bibitem[Cox (1963)]{ }
{Cox, J.P.} 1963, \textit{ApJ}, 138, 487

\bibitem[Christensen-Dalsgaard (2000)]{ }
{Christensen-Dalsgaard, J.} 2000, \textit{ASP-CS}, 210, 454

\bibitem[Degroote (2013)]{ }
{Degroote, P.} 2013, \textit{MNRAS}, 431, 255

\bibitem[Degroote et al.~(2010)]{ }
{Degroote, P., Briquet, M., Auvergne, M., et al.} 2010, \textit{A\&A}, 519, A38

\bibitem[Eddington (1919)]{ }
{Eddington, A.S.} 1919, \textit{MNRAS}, 79, 177

 \bibitem[Garc\'{i}a Hern\'{a}ndez (2009)]{ }
{Garc\'{\i}a Hern\'{a}ndez, A., Moya, A., Michel, E., et al.} 2009, \textit{A\&A}, 506, 79
 
\bibitem[Grigahc\`{e}ne et al.~(2010) ]{ }
{Grigahc\`{e}ne, A., Antoci, V., Balona, L.A., et al.} 2010, \textit{ApJ}, 713, L192

\bibitem[Guzik et al.~(2000) ]{ }	
{Guzik, J.A., Kaye, A.B., Bradley, P.A., Cox, A.N., \& Neuforge, C.} 2000, \textit{ApJ}, 542, L57

\bibitem[Gray \& Nagel (1989)]{ }
{Gray, D.F., \& Nagel, T.} 1989, \textit{ApJ}, 341, 421

\bibitem[Handler et al.~(2009)]{}
{Handler, G., Matthews, J.M., Eaton, J.A., et al.} 2009, \textit{ApJ}, 698, 56

\bibitem[Houdek et al.~(1999)]{ }
{Houdek, G., Balmforth, N.J., Christensen-Dalsgaard, J., \& Gough, D.O.} 1999, \textit{A\&A}, 351, 582

\bibitem[Huat et al.~(2009)]{ }
{Huat, A.-L., Hubert, A.-M., Baudin, F., et al.} 2009, \textit{A\&A}, 506, 95

\bibitem[Huber et al.~(2011)]{ }
{Huber, D., Bedding, T.~R., Stello, D., et al.} 2011, \textit{ApJ}, 743, 143

\bibitem[Kallinger \& Matthews (2010) ]{ }
{Kallinger, T., \& Matthews, J.M.} 2010, \textit{ApJ}, 711, L35

\bibitem[Kjeldsen \& Bedding (1995)]{ }
{Kjeldsen, H., \& Bedding, T.R.} 1995, \textit{A\&A}, 293, 87 

\bibitem[Kurtz (1982) ]{ }
{Kurtz, D.W.}  1982, \textit{MNRAS}, 200, 807
  
\bibitem[Landstreet et al.~(2009) ]{ }
{Landstreet, J.D., Kupka, F., Ford, H.A., et al.} 2009, \textit{A\&A}, 503, 973

\bibitem[Mathis et al.~(2013)]
Mathis, S., Neiner, C. \& Tran Minh, N., 2013, \textit{A\&A}, in press 

\bibitem[Neiner et al.~(2012)]{ }
{Neiner, C., Floquet, M., Samadi, R., et al.} 2012, \textit{A\&A}, 546, A47

\bibitem[Neiner et al.~(2013)]{ }
{Neiner, C., et al.} 2013,  H. Shibahashi and A. E. Lynas-Gray (eds.), \textit{Progress in Physics of the Sun and Stars: A New Era in Helio- and Asteroseismology"}, ASP-CS, in press


\bibitem[Pamyatnykh (2000)]{ }
{Pamyatnykh, A.A.} 2000, \textit{ASP-CS}, 210, 215

\bibitem[Poretti et al.~(2009)]{ }
{Poretti, E., Michel, E., Garrido, R., et al.} 2009, \textit{A\&A}, 506, 85

\bibitem[Porter \& Rivinius (2003)]{ }
{Porter, J.M., \& Rivinius, T.} 2003, \textit{PASP}, 115, 1153

\bibitem[Rogers et al.~(2012)]{ }
{Rogers, T.M., Lin, D.N.C., \& Lau, H.H.B.} 2012, \textit{ApJ}, 758, 6

\bibitem[Royer (2009)]{ }
{Royer, F.} 2009, \textit{Lecture Notes in Physics}, 765, 207

\bibitem[Samadi et al.~(2002) ]{ }
{Samadi, R., Goupil, M.-J., \& Houdek, G.} 2002, \textit{A\&A}, 395, 563

\bibitem[Samadi et al.~(2010)]{ }
{Samadi, R., Belkacem, K., Goupil, M.-J., Dupret, M.-A., Brun, A.S., \& Noels, A.} 2010, \textit{Ap\&SS}, 328, 253

\bibitem[Shiode et al.~(2013)]{}
Shiode, J.H., Quataert, E., Cantiello, M., Bildsten, L. 2013,  \textit{MNRAS}, 430, 1736


\bibitem[Simon et al.~(2002)]{ }
{Simon, T., Ayres, T.R., Redfield, S., \& Linsky, J.L.} 2002, \textit{ApJ}, 579, 800

\bibitem[Th\'eado et al.~(2012)]{ }
{Th\'{e}ado , S., Alecian, G., LeBlanc, F., \& Vauclair, S.} 2012, \textit{A\&A}, 546, A100

\bibitem[Tkachenko et al.~(2012)]{ }
{Tkachenko, A., Lehmann, H., Smalley, B., Debosscher, J., \& Aerts, C.} 2012, \textit{MNRAS}, 422, 2960

\bibitem[Uytterhoeven ey al.~(2011)]{ }
{Uytterhoeven, K., Moya, A., Grigahc\`ene, A., et al.} 2011, \textit{A\&A}, 534, A125 

\end{thebibliography}
\end{document}